\documentclass[11pt]{article}
\usepackage{amsfonts}
\usepackage{amssymb}
\usepackage{mathrsfs}
\usepackage{graphicx}
\usepackage{amsmath,amsthm,amssymb,amscd}
\usepackage[all]{xy}
\usepackage{subfig}
\usepackage{hyperref}
\usepackage{multirow}
\usepackage{color}
\usepackage{amsfonts}
\usepackage{float}
\newcommand{\cj}{{\cal J}}

\newcommand{\co}{{\cal O}}

\newcommand{\nn}{\nonumber}
\def\eqa{\begin{eqnarray}}
\def\eqae{\end{eqnarray}}
\def\eq{\begin{equation}}
\def\eqe{\end{equation}}
\def\be{\begin{equation}}
\def\ee{\end{equation}}
\def\bea{\begin{eqnarray}}
\def\eea{\end{eqnarray}}
\def\ba{\begin{array}}
\def\ea{\end{array}}
\def\bd{\begin{displaymath}}
\def\ed{\end{displaymath}}

\def\>{\rangle}
\def\<{\langle}
\def\a{\alpha}
\def\b{\beta}

\def\del{\delta}
\def\e{\epsilon}
\def\f{\phi}
  
\def\g{\gamma}
\def\h{\eta}

\def\m{\mu}
\def\n{\nu}

\def\s{\sigma}

\def\L{\Lambda}

\def\X{\Xi}

\def\pa{\partial}

\def\ul{\underline}

\newcommand{\fn}{\footnotemark\footnotetext}

\numberwithin{equation}{section}

\begin{document}

\begin{titlepage}
\hfill MCTP-13-03

\hfill NSF-KITP-13-039

\begin{center}
{\Large \bf  Toward a Higher-Spin Dual of Interacting Field Theories}\\
\end{center}

\vskip .7 cm

\vskip 1 cm
\begin{center}
{ \large Leopoldo A. Pando Zayas${}^a$ and Cheng Peng ${}^{a,b}$}
\end{center}

\vskip .4cm \centerline{\it ${}^a$ Michigan Center for Theoretical
Physics}
\centerline{ \it Randall Laboratory of Physics, The University of
Michigan}
\centerline{\it Ann Arbor, MI 48109-1120}
\bigskip\bigskip
\centerline{\em ${}^b$ Kavli Institute for Theoretical Physics}
\centerline{\em University of California}
\centerline{\em Santa Barbara, CA 93106-4030}\bigskip
\bigskip\bigskip

\vskip 1 cm

\vskip 1.5 cm
\begin{abstract}
We show explicitly how the exact renormalization group equation of interacting vector models in the large $N$ limit can be mapped into certain higher-spin equations of motion.
The equations of motion are generalized to incorporate a multiparticle extension of the higher-spin algebra, which reflects the ``multitrace" nature of the interactions in the dual field theory from the holographic point of view.
\end{abstract}

\end{titlepage}

\tableofcontents

\section{Introduction}
The holographic principle is  one of the fundamental ideas of modern physics; its explicit realization in the context of the Maldacena  conjecture \cite{Maldacena:1997re,Gubser:1998bc,Witten:1998qj} stands as a magnificent achievement. A natural limit of it is usually considered where a weakly coupled (super)gravity theory in asymptotically AdS spacetime is conjectured to be dual to a strongly interacting field theory. Working in this limit ultimately brought about potential applications to strongly coupled particle physics and, more recently, condensed matter physics. There are also other realizations of the holographic principle, one particularly interesting example is the duality between vector models and higher-spin theories \cite{Sundborg:2000wp,Mikhailov:2002bp,Sezgin:2002rt,Klebanov:2002ja} (see also \cite{Chang:2012kt,Aharony:2012nh,Aharony:2012ns,Giombi:2012ms} for further developments). One of the appealing features of the higher-spin/vector model duality is that it has a convenient weak coupling limit in the field theory side, namely the vector models with weak interactions. On the gravity side, the higher-spin theory in the bulk is conjectured to be related to the tensionless limit of string theory \cite{Sezgin:2002rt,HaggiMani:2000ru,Beisert:2004di}. This limit can also be understood as the ultra-high energy limit of string theory \cite{Gross:1987ar, Gross:1988ue} where stringy effects are dominant. Thus, the higher-spin/vector model duality could provide tools to shed some light on quantum gravity with strong interactions in the bulk AdS spacetime by studying relatively well understood weakly coupled field theories.

Many attempts to understand the physical origin of higher-spin holography, such as the bi-local field method \cite{Das:2003vw,Koch:2010cy,Jevicki:2011ss,Jevicki:2011aa,
deMelloKoch:2012vc}, the unfolding method \cite{Vasiliev:2012vf} and the ambient space approach \cite{Bekaert:2012vt}, have been pursued. A completely different line of attack was put forward by Douglas, Mazzucato and Razamat (DMR) \cite{Douglas:2010rc}. They considered free vector models with an effective nonlocal 2-point ``interaction", which encodes all the conserved currents with chemical potentials turned on. They then explicitly mapped the exact renormalization group (ERG) equation of this effective coupling constant to equations of motion in a higher-spin theory. The work \cite{Douglas:2010rc} is one realization of the general idea that reformulates the renormalization scale as an extra direction. Other realizations of this idea have been pursued \cite{Akhmedov:1998vf,deBoer:1999xf,Akhmedov:2002gq,Lee:2009ij,Heemskerk:2010hk,Lee:2010ub}.

In this paper we build on the approach of \cite{Douglas:2010rc} and show that their techniques can be suitably extended to incorporate interactions in the vector models. We consider interacting vector models obtained by deforming the free theory discussed in \cite{Douglas:2010rc} with irrelevant interactions (with $N$ dependence factored out explicitly), which makes sure that the RG flow goes back to the free theory.\fn{We need the RG flow to stay in the vicinity of the free theory fixed point so that the mapping from it to the bulk can be interpreted as the ``origin" of the duality between the free vector model and a bulk higher-spin theory. We will explain this in detail in the following sections.} We then explicitly map the ERG equations of the interactions close to the free theory fixed point to the equations of motion in a DMR type higher-spin theory \cite{Douglas:2010rc}. A key insight in our construction is provided by recent work of Vasiliev where the r\^ole of multi-particle states in  higher-spin theories is beginning to be elucidated \cite{Vasiliev:2012tv} (see also \cite{Vasiliev:1995dn,Vasiliev:1999ba,Bekaert:2005vh} for reviews of Vasiliev's  higher-spin theories). It is worth highlighting that the higher-spin theory discussed in \cite{Douglas:2010rc} is not explicitly of Vasiliev's type. However,  the method to construct multiparticle higher-spin algebras in \cite{Vasiliev:2012tv} is valid in general and can still be applied in our current discussion.

Unlike in the free theory case, our map is not exact: there are terms in the higher-spin equations of motion that do not appear in the ERG of the boundary field theory. These terms are suppressed by powers of $1/N$ and are proportional to the interactions on the boundary. Therefore our construction manifestly shows the breaking of the higher-spin symmetry by the appearance of interactions in the boundary field theory, which is consistent with the restrictions on (exact) higher-spin symmetries recently discussed in  \cite{Maldacena:2011jn,Maldacena:2012sf}.

The paper is organized as follows. In section \ref{ERG-to-HS:Free} we show the map from the renormalization group equation to the higher-spin equation of motion for a free theory while introducing necessary notation. This section is largely a review of \cite{Douglas:2010rc}. Section \ref{InteractingON} describes the interacting vector model in $D\geq 3$ dimensions using the exact renormalization group equation. In particular, we show how the multiparticle extension of the higher-spin algebra precisely accommodates the terms corresponding to some interactions in the vector model. We further discuss how to extend the previous map to the general correspondence between the ERG equation of interacting vector models and the equation of motion satisfying multiparticle extended higher-spin algebra. We conclude in section \ref{Conclusions}.

\section{From ERG to higher-spin: The free vector models}\label{ERG-to-HS:Free}
In this section, we briefly review the work of \cite{Douglas:2010rc} which consists in casting the ERG equations of a free vector model in a form that coincides with higher-spin equations of motion.
\subsection{The ERG equation}
A fundamental property of any quantum field theory is its behavior under the renormalization group (RG). In this work we focus on a particular realization of the RG known as the Exact Renormalization Group (ERG) developed originally by Wilson \cite{Wilson:1973jj}  (see section \S 11). It was fleshed out and formalized by Polchinski \cite{Polchinski:1983gv} whose presentation we shall follow (see also a recent review \cite{Rosten:2010vm}). For a theory living on $D$-dimensional spacetime with a generic interaction $S_{\rm int}$ in the UV, the ERG equation reads \fn{Here we assume that the interaction is semilocal  in the sense that momentum dependent coupling constants admit Taylor expansions in  momentum space.}:
\begin{equation}\label{erg}
  d_\L S_{\rm int}=-\int d^{(D)}p \frac{d_{\L} K(p^2/\L^2,\vec{e})}{K(p^2/\L^2,\vec{e})}\bigg(1+\frac{K(p^2/\L^2,\vec{e})}{p^2}\big(\frac{\pa S_{\rm int}}{\pa \bar{\f}(p)}\frac{\pa S_{\rm int}}{\pa {\f}(p)}+\frac{\pa^2 S_{\rm int}}{\pa \bar{\f}(p) \pa {\f}(p)}\big)\bigg)\,,
\end{equation}
where $d_\Lambda = \Lambda \partial/\partial\Lambda$ and we have separated out the kinematic part $S_{\rm free}$ from the action. We have introduced a momentum cutoff function $K(p^2/\L^2,\vec{e})$ such that the propagator vanishes quickly when $p^2>\L^2$ and goes to $1/p^2$ for $p^2<\L^2$. Equation \eqref{erg} is derived by requiring the partition function of the (effective) field theory to be invariant under the change of the physical scale, $\Lambda$. \fn{We have explicitly allowed a normalizing field-independent term left out in \cite{Polchinski:1983gv}, this term accounts for the factor of $-1$ in the above form of the ERG equation.}

In addition, we have chosen an explicit reference point, $\vec{e}=\{e^0,\ldots,e^{D-1}\}$, in the D-dimensional spacetime and the momentum cutoff function depends on the position of this reference point. We require that the low energy effective theory does not depend on
 the choice of this reference point and that implies the invariance
 under any spacetime translation.\fn{Choosing a reference/marked point in spacetime is conceptually analogous to choosing a point $\L$ in the direction of energy scale. The invariance of the partition functions under any change of this reference point is merely a reflection that the quantum field theory is defined on an affine space: choosing a reference point is the same as adding an origin to the affine space to identify it with the underlying vector space, and except for this the origin is not a special point in any sense.}
This leads to another set of equations constraining the changes of the effective action in response to a change of the vector $\vec{e}$
\begin{equation}\label{ergspace}
  d_{e^a} S_{\rm int}=-\int d^{(D)}x \frac{d_{e^a} K(\square/\L^2,\vec{e})}{K(\square/\L^2,\vec{e})}\bigg(1+\frac{K(\square/\L^2,\vec{e})}{\square}\big(\frac{\pa S_{\rm int}}{\pa \bar{\f}(x)}\frac{\pa S_{\rm int}}{\pa {\f}(x)}+\frac{\pa^2 S_{\rm int}}{\pa \bar{\f}(x) \pa {\f}(x)}\big)\bigg)\,,
\end{equation}
where $\square\equiv \pa_b \pa^b$, $a,b=0,\ldots,D-1$ and we have expressed the ERG equation in position space for later convenience.
This set of equations is derived exactly as the derivation of \eqref{erg}, which was carried out in \cite{Polchinski:1983gv}. A direct consequence of this fact is that \eqref{ergspace} is almost identical to \eqref{erg} only with $d_{\L} K(p^2/\L^2,\vec{e})$ replaced by $d_{e^i} K(p^2/\L^2,\vec{e})$ in the momentum space.

Following DMR \cite{Douglas:2010rc}, one considers the ERG equation for the theory of $N$ free complex scalar fields $\phi^A(x)$ in $D$ dimensions
\be\label{sintspace}
S=\sum\limits_A\int d^Dx|\partial\phi^A(x)|^2 -\int d^D x d^D y B(x,y) \bar{\phi}^A(x)\phi^A(y).
\ee

In momentum space, the action takes the form
\be
S=\int d^Dp \, d^Dq \left(P(p,q)-B(p,q)\right)\bar{\phi}^A(q)\phi^A(p)\,,
\ee
where
\bea
P(p,q)&=& p^2 K^{-1}(p^2/\L^2)\delta^{(D)}(p-q), \qquad \nn  B(p,q)=\int \frac{d^Dx\, d^Dy}{(2\pi)^{2D}}e^{-iqx+ipy}B(x,y)\\
 &&\phi^A(x)=\int \frac{d^Dp}{(2\pi)^D} e^{ipx}\phi^A(p)\,.\label{phift}
\eea

For the quadratic action above we identify the ``interacting'' piece as:
\be
S_{int}=\int d^Dp d^Dq B(p,q)\bar{\phi}^A(q)\phi^A(p).
\ee
As in \cite{Douglas:2010rc}, this choice of bi-local interaction is a collective notation for operators of the form $J_s \sim \phi^A \partial_{\mu_1}\ldots \partial_{\mu_s}\phi^A$, that is, the set of singlet conserved currents in the free field theory.
Substituting into the ERG  and considering the coefficient of the $\bar \phi \phi $ term we obtain the equation
\be\label{b1run}
\L\frac{\partial B(p,q)}{\partial \L}=-
\int \frac{d^Ds}{s^2}\L \frac{\partial K(s^2/\L^2,\vec{e})}{\partial \L}B(s,q)B(p,s).
\ee

In order to cast the ERG equations in the language of higher-spin theories, it is convenient to introduce the following definitions
\bea
&&\a_r=\frac{d_\L K(p^2/\L^2,\vec{e})}{p^2}\del^{(D)}(p-q)\,,\\
&&B(p/\L,q/\L)= \L^{2-D-|s|-|t|}\,B_{\underline{st}}\,p^{\underline{s}}\,q^{\underline{t}}\ ,\label{bexpansion}\\
&&\a_r^{\underline{st}} = \L^{2-D-|s|-|t|}\,\int d^Dp \int d^Dq\ \a_r(p,q) p^{\underline{s}}\,q^{\underline{t}}\,,
\eea
where $\ul{s}$ stands for an array of indices taking values in the set $a=0,\ldots,D-1$, and $|s|>0$ represents the length of the array $\ul{s}$.\fn{ We impose the condition that $|s|\neq 0$ for the following reason. Suppose in the expansion \eqref{bexpansion} we have a term with $|s|=0$, which means we have a contribution $\sum_t B_{\ul{t}} q^{\ul{t}}$ to the $B(p/\L,q/\L)$. We can inverse Fourier transform this contribution to configuration space and that will give a term $\int d^D y B(0,y)\bar\f^A(0)\f^A(y)\equiv \int d^D y \cj^A(y)\f^A(y)$. In this work we will turn off any source of a single field $\f^A(x)$ or $\bar\f^A(x)$ which has a free $U(N)$ indices, since it may lead to the breaking of the $U(N)$ symmetry. This requires that in the expansion \eqref{bexpansion} neither $|s|$ nor $|t|$ can be zero.} Then the ERG equation takes an instructive form
\be\label{ergfnl}
\frac{d}{d\L} B_{\underline{st}}=-B_{\underline{si}}\,\a_r^{\underline{ij}}\,B_{\underline{jt}}+\L^{-1}(|s|+|t|+D-2)\,B_{\underline{st}}\,.
\ee

Similar equations can be derived from \eqref{ergspace}. Plugging the interacting part of \eqref{sintspace} into \eqref{ergspace} we get
\be\label{b1runspace1}
 d_{e^a} \big(B(x,y)\bar{\f}^A(x)\f^A(y)\big)=-\int \frac{d^D z}{\square} d_{e^a}{K(\square/\L^2,\vec{e})}B(x,z)B(z,y)\bar{\f}^A(x)\f^A(y)\,.
\ee
Since \eqref{ergspace} is derived by requiring invariance of the partition function under a change of the reference point $\vec{e}$, the meaning of the left hand side of \eqref{b1runspace1} becomes transparent only after we consider the change of the point $\vec{e}$. To do so, we use an equivalent (passive) realization of the change of $\vec{e}$, namely a rigid translation of the coordinates $\vec{x}\to \vec{x}+\vec{e}$. The invariance of this transformation means that we can rewrite \eqref{b1runspace1} as
\begin{eqnarray}
\nn &&  \hspace{-6mm}d_{e^a} \big(B(x+e,y+e)\bar{\f}^A(x+e)\f^A(y+e)\big)  = \\
&& \hspace{-6mm}\qquad -\int \frac{d^D z}{\square} d_{e^a}{K(\square/\L^2,\vec{e})}B(x+e,z+e)B(z+e,y+e)\bar{\f}^A(x+e)\f^A(y+e)\label{space1}
\end{eqnarray}
To pull the $\bar{\f}^A(x+e)\f^A(y+e)$ out of the derivative, we use the Fourier representation of the fields \eqref{phift}, which leads\fn{Notice that here we have used the fact that the translation $\vec{x}\to \vec{x}+\vec{e}$ is rigid, which means $\vec{e}$ does not depend on $x$. This makes sure that the Fourier transformation \eqref{space2} is not messed up in the presence of $\vec{e}$. Moreover, the requirement that $\vec{e}$ does not depend on $x$ is also used in the derivation of \eqref{ergspace}, which guarantees that the $d_{e^a}$ commutes with spacetime derivatives.}
\begin{eqnarray}
\nn && \hspace{-16mm} d_{e^a} \big(B(x+e,y+e)\bar{\f}^A(x+e)\f^A(y+e)\big)\\
\nn &=& d_{e^a} \big(B(x+e,y+e)\int \frac{d^Dp}{(2\pi)^D} e^{-ip(x+e)}\bar{\f}^A(p)\frac{d^D q}{(2\pi)^D} e^{iq(y+e)}\bar{\f}(q)\big) \\
   &=&  \big(d_{e^a}B(x+e,y+e)+i(q_a-p_a)B(x+e,y+e)\big)\bar{\f}^A(x+e)\f^A(y+e)\,.\label{space2}
\end{eqnarray}
Plugging this back to \eqref{space1}, rewriting $x+e,\,y+e,\,z+e$  as $x,\,y,\,z$ for simplicity, we have
\begin{eqnarray}
\nn  &&\hspace{-8mm}d_{e^a}B(x,y)+i(q_a-p_a)B(x,y) \\
   &=&  -\int \frac{d^D z}{\square} d_{e^a}{K(\square/\L^2,\vec{e})}B(x,z)B(z,y),
\end{eqnarray}
where the momenta $p,q$ should be understood in terms of \eqref{space2}.

It is more convenient to express the above result in the momentum space,
\begin{eqnarray}
  &&\hspace{-8mm}d_{e^a}B(p,q)= -\int \frac{d^D s}{s^2} d_{e^a}{K(s^2/\L^2,\vec{e})}B(p,s)B(s,q)-i(q_a-p_a)B(p,q)\,.
\end{eqnarray}
We can then do an expansion in momentum space as in \eqref{ergfnl}, which gives
\be\label{Bifinal}
d_{e^a}\,B_{\underline{st}}=-B_{\underline{sx}}\,\a_i^{\underline{xy}}\,B_{\underline{yt}}+i
(p^a-q^a)\, B_{\underline{st}}.
\ee
where $d_{e^{a}}=e^{a}\partial/\partial e^{a}$. Note that the second term in the right hand side of equation  \eqref{Bifinal} is proportional to momentum conservation as should be the case for interactions that are invariant under $\vec{x}\to \vec{x}+\vec{e}$.

\subsection{The star product}\label{starsec}

Higher-spin algebras are conveniently defined with the help of a star product. Thus, in order to map the ERG equations derived in the previous section, we first introduce the star product that will be a key ingredient in the following analysis. Following \cite{Vasiliev:1999ba, Douglas:2010rc}, we introduce auxiliary variables $y^\a,\bar{y}_\a,z^\a,\bar{z}_\a$, with the indices $\a\in \{\bullet,r,0,\ldots, D-1\}$. To simplify our computation, we adopt two equivalent expressions for the auxiliary variables:
\begin{equation}\label{twistereqn}
y^\a\equiv y^{1,\a}\,, \quad \bar y_\a\equiv y^{2,\a}\,, \quad z^\a\equiv z^{1,\a}\,, \quad \bar z_\a\equiv z^{2,\a}\,.
\end{equation}

In terms of these variables, the star product admits a differential expression
\begin{equation}\label{diffstarplus}
  (f*g)(y,z)=
   f(y,z)\exp\bigg(-\frac{1}{2}\h_{\a\b}\e_{ij}(\frac{\overleftarrow{\pa}}{\pa z^{i,\a}}+\frac{\overleftarrow{\pa}}{\pa y^{i,\a}})
   (\frac{\overrightarrow{\pa}}{\pa y^{j,\b}}-\frac{\overrightarrow{\pa}}{\pa z^{j,\b}})\bigg)g(y,z)\,,
\end{equation}
where
the index $\a,\b\in \{\bullet,r,0,\ldots, D-1\}$ and $i,j=1,2$ label the (un)barred variables. The extended metric is taken to be $\h_{\a\b}=diag{(-1,1,\h^{\rm CFT}_{aa})}$, where $a,b=0,\ldots,D-1$ label the (boundary) spacetime.

For later convenience, we list some facts about the star product \eqref{diffstarplus}. First, it is useful to define new variables $Y,\,Z,\, \bar{Y},\,\bar{Z}$ via
 \be
y^a=Y^a+ Z^a,\qquad
\bar y_a = \frac{1}{2}(\bar Y_a- \bar Z_a),\qquad z^a=Z^a- Y^a,\qquad
\bar z_a = \frac{1}{2}(\bar Y_a+ \bar Z_a)\,.
\ee
The star product \eqref{diffstarplus} in terms of these $Y,\, Z$ variables reads
\begin{equation}\label{diffstarYZ}
 (f'*g')(Y, Z,\bar{Y},\bar{Z})=
  f'(Y,\bar Y,Z, \bar Z)\exp\bigg(
  \frac{\overleftarrow{\pa}}{\pa \bar Y_{a}}\frac{\overrightarrow{\pa}}{\pa Y^{a}}+
  \frac{\overleftarrow{\pa}}{\pa Z^{a}}\frac{\overrightarrow{\pa}}{\pa \bar Z_{a}})
  \bigg)g'(Y,\bar Y,Z, \bar Z)\,.
\end{equation}

With this definition, it is convenient to construct a
Klein-like operator
\begin{equation}\label{idnull}
  G=\exp(-Y^a\bar Y_a-Z^a \bar Z_a)\,.
\end{equation}
One can then check that
\begin{eqnarray}
  \bar Y *G = 0 \,,  \quad  Z*G = 0\,, \quad  G*Y = 0\,, \quad   G* \bar Z=0 \,. \label{mprodv}
\end{eqnarray}
However, the order of operators in these simple vanishing relations is crucial, if we flip the order, then the star products are no longer vanishing, namely
 \begin{eqnarray}
   Y *G = Y G \,, \quad \bar Z*G = \bar Z G\,,\quad  G*\bar Y = G \bar Y\,, \quad   G*  Z= G Z\,, \label{mprodnv}
\end{eqnarray}
where the right hand side are usual products. Another useful relation in the computation reads
\begin{equation}\label{mmid}
  G*G=GG+G(Y \bar Y+Z \bar Z)G+\frac{1}{2}G(Y \bar Y+Z \bar Z)^2G+\ldots=GG^{-1}G=G.
\end{equation}

 In fact, we can expand any function $f(Y,Z,\bar Y,\bar Z)$ that involves the operator $G$ and admits a formal power expansion in terms of $Y,Z,\bar Y,\bar Z$ into the form $$f(Y,Z,\bar Y,\bar Z)=\sum\limits_{\underline{a},\underline{b},\underline{c},\underline{d}}f^{\underline{abcd}}Y^{\underline{a} } * \bar  Z^{\underline{b}} * G * \bar Y^{\underline{c}} *Z^{\underline{d}}\,.$$
 The above presentation follows from the properties \eqref{mprodnv}: each term in this formal expansion can be written as a product of usual multiplications between commuting variables $Y^{\underline{a} } \bar  Z^{\underline{b}} \bar Y^{\underline{c}} Z^{\underline{d}}G=Y^{\underline{a} } * \bar  Z^{\underline{b}} * G * \bar Y^{\underline{c}} *Z^{\underline{d}}$.
 The star product of two terms in the above formal expansion has the following simple form
 \begin{eqnarray}
      \nn  && (Y^{\underline{a} } * \bar  Z^{\underline{b}} * G * \bar Y^{\underline{c}} *Z^{\underline{d}})*(Y^{\underline{a'} } * \bar  Z^{\underline{b'}} * G * \bar Y^{\underline{c'}} *Z^{\underline{d'}}) \\
       && \qquad =|c|!|d|!\del_{|c|,|a'|}\del_{|d|,|b'|}\del^{c_1}_{(a'_1}\ldots\del^{c_n}_{a'_n)}\del^{d_1}_{(b'_1}\ldots\del^{d_n}_{b'_n)}
      Y^{\underline{a} } * \bar  Z^{\underline{b}} * G* \bar Y^{\underline{c'}} *Z^{\underline{d'}}\,.\label{starcontraction}
 \end{eqnarray}

\subsection{Mapping ERG to higher-spin equations of motion}\label{mapfreesec}
With the star product defined above, it is possible to connect the ERG equations \eqref{ergfnl} and \eqref{Bifinal} with  higher-spin equations of motion in the AdS$_{D+1}$ spacetime background.

We start constructing the bulk higher-spin theory by introducing the following gauge connections that describe the AdS background \cite{Douglas:2010rc}
\be\label{adsconnection}
W^{(0)}_\mu = \frac{1}{r}\, P_\mu\,,
\ee
where
\bea\label{adscmp}
P_{r}&=&\bar z_{r} z^{\bullet}-\bar z_{\bullet}z^{r},\\
P_a&=&\bar z_a\,(z^{\bullet}-z^{r})-(\bar z_{\bullet}-\bar z_{r})\,z^a\,.\nonumber
\eea
The connection satisfies the flatness condition evaluated using the star product \label{diffstar}
\be\label{flatads}
dW^{(0)}+W^{(0)}*W^{(0)}=0\,.
\ee

The higher-spin fields are treated as fluctuations around this AdS background \eqref{adsconnection}. This idea is implemented in \cite{Douglas:2010rc} by defining $W=W^{(0)}+\delta W$ with $\delta W_r =r B*\a_r$,\fn{Here we include an extra factor of $r$ to correct the $r$ counting.} and
\bea
&&B(y,z,\bar y,\bar z)= i^{|s|-|t|}r^{D-2}\,B_{\underline{st}}\,Y^{\underline{s}} \,Z^{\underline{t}}\, e^{-Y\bar Y-Z\bar Z}\,(\bar z_{r}-\bar z_{\bullet})^{|s|+|t|}\,,\label{defBtwt}\\
&&\a_r(y,z,\bar y,\bar z)=-\frac{(-i)^{|t|-|s|}}{|s|!|t|!} r^{-D}\,\a_r^{\underline{st}}\,\bar Y_{\underline{t}} \,\bar Z_{\underline{s}}\, e^{-Y\bar Y-Z\bar Z}\,(\bar z_{r}-\bar z_{\bullet})^{-|s|-|t|}\, .\label{defalptwt}
\eea
Then with the help of the properties in section \ref{starsec} it is easy  to show that
\bea
\nn \hspace{-10mm}\frac{d}{d r}\,B+W_r*\, B- B*\, \widetilde W_r&=&i^{s-t}r^{D-3}\,\,Y^{\underline{s}} \,Z^{\underline{t}}\, e^{-Y\bar Y-Z\bar Z}\,(\bar z_{r}-\bar z_{\bullet})^{|s|+|t|}\\
&&\times \bigg((D-2+(|s|+|t|)) B_{\underline{st}}+r \frac{\pa}{\pa r} B_{\underline{st}}-B_{\underline{su}}\,\a_\mu^{\underline{uq}}B_{\underline{qt}}\bigg)\,,\label{bwwtr}
\eea
where  $\widetilde{W}=W^{(0)}$. Notice that the terms in the big parenthesis in \eqref{bwwtr} coincide with the ERG equation \eqref{ergfnl}
with the identification $\L=\frac{1}{r}$ and hence $d_\L=\L\frac{\pa}{\pa \L}=-r \frac{\pa}{\pa r}$.

So the ERG equation \eqref{ergfnl} implies the following equation
\be\label{bwwt}
\frac{d}{dr}\,B+W_r*\, B- B*\, \widetilde W_r=0\,.
\ee
Equations in the other directions $x_a\neq r$ can be derived in a similar way. It is shown in \cite{Douglas:2010rc} that the $[W^{(0)},B]_*$ reproduces the linear term on the right hand side of \eqref{Bifinal}. The term $\del W_a*B=r\, B*\a_a*B$, where $\a_a$ is defined as
\begin{equation}\label{alpspace}
  \a_a(y,z,\bar y,\bar z)=-\frac{(-i)^{|t|-|s|}}{|s|!|t|!} r^{-D}\,\a_a^{\underline{st}}\,\bar Y_{\underline{t}} \,\bar Z_{\underline{s}}\, e^{-Y\bar Y-Z\bar Z}\,(\bar z_{r}-\bar z_{\bullet})^{-|s|-|t|}\, ,
\end{equation}
reproduces the nonlinear term on the right hand side of \eqref{Bifinal} in the same way as in \eqref{bwwtr}. This similarity between the computation in the $r$ direction and the $x_a$ direction is not surprising since the ERG equations \eqref{b1run} and the $x_a$ translation equation \eqref{b1runspace1} are identical provided that we replace $\a_r$ with $\a_a$. Taking the two pieces together, we have the following equation for $B$ in the $x_a$ direction
\be\label{bwwtmu}
\frac{d}{d x_a}\,B+W_a*\, B- B*\, \widetilde W_a=0\,,
\ee
whose vanishing is a direct consequence of \eqref{Bifinal}.

Further more, following \cite{Douglas:2010rc} we impose the covariance condition
 \be
d\a+W^{(0)}*\,\a+\a*\,W^{(0)}=0\,.
\ee
which should be understood as a constraint on the cutoff function $K(p^2/\L^2,\vec{e})$. Then right star multiplying \eqref{bwwt} and \eqref{bwwtmu} by $\a_\m$ and taking into account the flatness condition of the AdS background \eqref{flatads}, we arrive at the following equations which turn out to be a flatness conditions of the full connection
\be\label{fullflat}
dW+W * W=0\,,\quad d\widetilde W+\widetilde W * \widetilde W=0\,.
\ee
Thus, the ERG equations are equivalent to higher-spin equations of motion expressed as flatness conditions of the gauge connection. One can extract the equation of motions for gauge fields with different spins by an expansion of the field $W$ and $B$ in powers of the auxiliary variables.

\section{From ERG to higher-spin: The interacting vector models}\label{InteractingON}
Now we are ready to deform the free theory at UV with higher-point interactions. In the following discussion, we only add in irrelevant deformations. In D dimension, relevant deformations correspond to operators of the form $(\bar{\f}\f)^n$ with $n< \frac{D}{D-2}$. In this paper, we give a uniform treatment for CFT in $D\geq 3$ dimensional spacetime, so we only put in operators with $n> 3$ in our computation. One exception is that the operator with $n=3$ is irrelevant in $D>3$ and exact marginal in $D=3$ \cite{Aharony:2011jz}, so we will put it in our discussion as well. In summary, we consider operator deformations of the form $(\bar{\f}\f)^n$ with $n \geq 3$ in $D \geq 3$.

We do not include the relevant deformations since they will drive the theory to some non-trivial strongly coupled fixed point or some strongly coupled massive confining theories deep in the IR. In this work one of our motivations is the duality between a 3D free CFT on the boundary and a higher-spin theory in the bulk.\fn{There are other versions of holographic dualities involving higher-spin theories in the bulk with interacting CFT's on the boundary. The higher-spin/critical $O(N)$ vector model duality has been shown, in \cite{Hartman:2006dy, Giombi:2011ya}, to follow from the free CFT duality order by order in  $\frac{1}{N}$.} Thus we do not want our theory to flow to some strongly coupled regime, which indicates that we should only consider the (marginally) irrelevant deformations.

However, this is not enough to keep us close to the free theory fixed point. Note that one essential difference between the ERG in the free theory \cite{Douglas:2010rc} and in our interacting theory is that in the Wilsonian approach, all types of interactions, including the relevant interactions, are generated along the RG flow. These dynamically generated interactions will ultimately drive us to the stable Wilson-Fisher fixed point in the IR in $D<4$ dimension. So in order to stay in a vicinity of the free theory, we have to put in small irrelevant perturbations and run down infinitesimally from energy scale $\Lambda$ to $\Lambda-\epsilon$. Then the effects of the relevant operators are suppressed by $\co(\e)$ and we then have a valid approximation.

\subsection{The interacting ERG equation}

We consider a general UV action with nonlocal interactions
\begin{eqnarray}
  S_{\rm}\hspace{-3mm}&=&\hspace{-3mm}\int d^D p d^D q \big(P(p,q) -B^{(1)}(p,q) \big) \bar{\f}(q)^A\f(p)^A\label{sgdeformUV}\\
\nn  &&\hspace{-3mm}- \frac{1}{N^{n-1}}\int \bigg(\prod\limits_{j=1}^n d^D p_j d^D q_j \bar{\f}^{A_j}(q_j)\f^{A_j}(p_j)\bigg)\, B^{(n)}(p_1,\ldots,p_n,q_1,\ldots,q_n) \,.
\end{eqnarray}
As stated above the sum over repeated indices implies that we restrict our treatment to the singlet sector. Notice that we have explicitly written out a $1/N^{n-1}$ factor in the interaction term since the interaction we put in is a multitrace operator.

Using \eqref{erg}, we can compute the running of the coupling constants $B^{(1)}$ and $B^{(n)}$, namely the ERG, in the Wilsonian approach. The ERG of $B^{(1)}$ turns out to be the same as \eqref{b1run}. The ERG of the coupling constants $B^{(n)}$ reads\fn{Here we leave out the free energy part which is the term that does not depend on $(\bar{\f}(p)\f{(p)})^n$. Its form is exactly as discussed in \cite{Douglas:2010rc} and can be accounted for, if desired, by the appropriate field-independent transformation used in equation (\ref{erg}).}
\begin{eqnarray}
  \nn \hspace{-10mm}&&  \hspace{-12mm}d_\L B^{(n)}(p_1,\ldots,p_n,q_1,\ldots,q_n) =\\
  &&\hspace{-9mm}-\int d^D r\frac{d_\L K(r^2/\L^2,\vec{e})}{r^2}\sum\limits^n_{k=1} \bigg( B^{(n)}(\ul{p},\ul{q})\big|_{q_k=r}B^{(1)}(r,q_k)+ B^{(n)}(\ul{p},\ul{q})\big|_{p_k=r}B^{(1)}(p_k,r)\bigg)\,.\label{nptrun1}
\end{eqnarray}
We have an important comment on this equation. From \eqref{erg} we see that there will be extra terms with the schematic form $B^{(n+1)}+\sum_{l=2}^n B^{(l)}B^{(n-l+1)}$ that contribute to equation \eqref{nptrun1}: these terms will be generated along the RG flow even if they are not present in the initial deformed Lagrangian \eqref{sgdeformUV}. However, they are all of order $\co(\e)$ as we only run down infinitesimally in the energy scale, as we discussed at the beginning of this section. Therefore, their contributions to equation \eqref{nptrun1} drop out, which makes \eqref{nptrun1} a good approximation. A similar approximation is also used in the ERG of $B^{(1)}$.

We assume the coupling constants to be semi-local, which means we can expand the coupling constants in power series of  momenta \begin{eqnarray}
    B^{(n)}(\ul{p},\ul{q})\hspace{-3mm} &=&\hspace{-3mm}  \L^{n(2-D)-\sum_i(|\ul{s_i}|+|\ul{t_i}|)}B^{(n)}_{\ul{s_1},...\ul{s_n}, \ul{t_1},...\ul{t_n}}\prod\limits_{i=1}^n
  p_i^{\ul{s_i}}q_i^{\ul{t_i}}\,,\label{defbninerg}\\
\a_r^{\underline{st}}&=& \L^{2-D-|s|-|t|}\,\int d^Dp \int d^Dq\ \a(p,q) p^{\underline{s}}\,q^{\underline{t}},\,\label{defalpinerg}
\end{eqnarray}
where in \eqref{defbninerg} the factor ${n(2-D)}$ is the mass dimension of $B^{(n)}(\ul{p},\ul{q})$ in momentum space, this factor is included to make sure that $B^{(n)}_{\ul{s_1\ldots s_n t_1\ldots t_n}}$ is dimensionless. Note that $\a^{\ul{st}}$ has mass dimension $-1$. Plugging these into \eqref{nptrun1}, we have
\begin{eqnarray}
  d_\L B^{(n)}_{\ul{s_1},...\ul{s_n}, \ul{t_1},...\ul{t_n}} &=& -\L^{-1} (n(2-D)-\sum_i(|\ul{s_i}|+|\ul{t_i}|))B^{(n)}_{\ul{s_1},...\ul{s_n}, \ul{t_1},...\ul{t_n}}\label{nptrun}\\
\nn &&+\sum\limits^n_{k=1} \bigg( (B^{(n)}_{\ul{s_1},...\ul{s_n}, \ul{t_1},...\ul{t_n}}\big|_{\ul{t_k}=\ul{a}})\,\a^{\ul{a},\ul{b}}B^{(1)}_{\ul{b},t_k}+ B^{(1)}_{\ul{s_k},\ul{a}}\,\a^{\ul{a},\ul{b}}(B^{(n)}_{\ul{s_1},...\ul{s_n}, \ul{t_1},...\ul{t_n}}\big|_{\ul{s_k}=\ul{b}})\bigg)\,.
\end{eqnarray}
Thus, we see that the ERG equation for a general interaction gives rise to an infinite set of equations, most importantly,  this set of equations is characterized by strings of indices $\ul{s_1\ldots s_n t_1\ldots t_n}$, and on the right hand side, all possible terms with the free indices being $\ul{s_1\ldots s_n t_1\ldots t_n}$, whose relative order is fixed, are present with simple coefficients. This is the master equation that connects to equations of motion for higher-spin fields in the bulk.

\subsection{Multiparticle higher-spin algebra}\label{multisec}
As indicated in \eqref{sgdeformUV}, we have considered operator deformations that look like ``multitrace" operators. From the holographic point of view, these multitrace deformations should be related to the multiparticle states in the bulk. In the case we are interested in, this suggests that the bulk dual should contain fields that transform under some ``multiparticle" version of the higher-spin algebra. In \cite{Vasiliev:2012tv}, Vasiliev has constructed this algebra explicitly. Since the multiparticle extension of higher-spin algebras plays a crucial role in the following discussion, we will first briefly review one realization of it following \cite{Vasiliev:2012tv}, which turns out to be very convenient for us.

The generators of the (single particle) higher-spin algebra can be collectively denoted as $\X=(1, y^\m, z^\m, y^\m y^\n, \ldots )$. For $n$-particle algebra $M_n$, we consider $n$ species of these generators, which can be collectively denoted as $\X_i=(1, y_i^\m, z_i^\m, y_i^\m y_i^\n, \ldots )$.
The multiplication rules among them are defined as
\begin{equation}\label{flavorcmt}
  \X_i^\a * \X^\b_i =f_\g^{\a\b}\X^\g_i\,,\qquad \X^\a_i*\X^\b_j=\X^\b_j*\X^\a_i\,,
\end{equation}
where $i,j$ labels different sets of auxiliary variables and $\a,\b$ labels different basis in one set of the basis.
The multiparticle algebra can be realized as a subalgebra of the enveloping algebra of the algebra defined by \eqref{flavorcmt}, whose generators are  polynomials of $\X_i$ that are symmetric under the $S_n$ permutation group of different species.

With these generators in different species, we can define the following  functions that will be useful in the discussion below.

In the context of multiparticle algebra, the definition of the $\a_\m(y^a,z^a)\equiv \a_\m{(\X)}$ now depends on which set of $\X$ we use. It is natural to define a general $\a_\m(\X_j)$ function that is obtained from $\a_\m{(\X)}$ by identifying $\X$ with $\X_j$.

We can further define
\begin{eqnarray}
 &&\hspace{-10mm} C(\X_j,s_j,t_j)= i^{s_j-t_j}\,r^{D-2}\,Y_j^{\ul{s_j}}*G_j*Z_j^{\ul{t_j}}( {(\bar z_j)}_{r}- {(\bar z_j)}_{\bullet})^{s_j + t_j}\, ,\label{ckyz}\\ &&\hspace{-10mm} E(\X,s,t) =
 \a_\m(\X)* C(\X,s,t)=\sum\limits_p\bigg(-\frac{i^{p-t}}{p!}r^{-2}\,\a_\mu^{\underline{ps}}\,( \bar Z_{\underline{p}}\, *G\,*Z^{\underline{t}})\,(\bar z_{r}-\bar z_{\bullet})^{t-p}\bigg)\, ,\label{est}\\
 &&\hspace{-10mm} H(\X,s,t) =
  C(\X,s,t)*\a_\m(\X)=\sum\limits_q \bigg(-\frac{i^{s-q}}{q!}r^{-2}\,\a_\mu^{\underline{tq}}\,(Y^{\underline{s}} \,*G\, *\bar Y_{\underline{q}})\,(\bar z_{r}-\bar z_{\bullet})^{s-q}\bigg)\,,\label{hst}
\end{eqnarray}
where we have suppressed the species indices $j$ in \eqref{est} and \eqref{hst} and the $\X$'s therein are of the same specie.
From \eqref{flavorcmt}, we have the following relations
\begin{equation}\label{cidty}
  C(\X_l,s_l,t_l)*C(\X_l,s_l,t_l)=0\,,\qquad C(\X_j,s_j,t_j)*C(\X_l,s_l,t_l)=C(\X_l,s_l,t_l)*C(\X_j,s_j,t_j)\,, \quad j\neq l\,.
\end{equation}
The proof of these relations is simple. The first half is a direct consequence of \eqref{starcontraction} since we always have $G*Y$ in the product. The second half is a direct consequence of \eqref{flavorcmt}.
Another relation we need is
\begin{eqnarray}
\nn && H(\X_j,s_j,t_j)*C(\X_j,p_j,q_j)
= -r^{-2 }\a_\mu^{\underline{tp}}\, C(\X_j,s_j,q_j)=C(\X_j,p_j,q_j)*E(\X_j,s_j,t_j)\,.
\end{eqnarray}

\subsection{Mapping to higher-spin equation of motions}
We connect the multiparticle algebra with the ERG equation \eqref{nptrun} via the follow schematic map
 \begin{equation}\label{mapmomtwtmtp}
  p_j^a \,\to\, i\frac{Y_j^a}{r}\,,\qquad q_j^a \, \to\, -i\frac{Z_j^a}{r}\,.
\end{equation}
Note in \eqref{nptrun} the monenta carry ``boundary" indices with $a\in \{0,\ldots, D-1\}$. To carry out the star products, we first double the $Y$ and $Z$ auxiliary fields in \eqref{mapmomtwtmtp} to the set of $\{Y^a,Z^a,\bar{Y}^a,\bar{Z}^a\}$, and we will do a truncation by setting
\begin{equation}\label{projection}
\bar{Y}^a=0\,,\qquad \text{and}\qquad \bar{Z}^a=0 \,,
\end{equation}
at the very end of the computation since the physical information is all encoded in the $Y,Z$ part.\fn{Notice that this projection is in the same fashion as in Vasiliev higher-spin theory, where the physical information is encoded in the terms that only depend on $Y$ and $\bar{Y}$ variables. To extract the physical information, we need to set $Z=0$ and $\bar{Z}=0$ at the end of the computation in the Vasiliev higher-spin theory (e.g. \cite{Giombi:2012ms}). In our higher-spin theory, the physical information is encoded in $Y,Z$ \eqref{mapmomtwtmtp}, so a similar projection in our case is given by \eqref{projection}.}
The mapping \eqref{mapmomtwtmtp} suggests the following definition
\begin{equation}\label{defbntwt}
 B^{(n)}(\X_{l_1},\ldots,\X_{l_n}) = B^{(n)}_{\ul{s_1,\cdots,s_n t_1 ,\cdots, t_n}} \prod\limits_{j=1}^n C(\X_{l_j},s_j,t_j)\,,
\end{equation}
where repeated indices of $s_k, t_k$ are summed over implicitly.

Following the properties in section \eqref{multisec} with $1\leq k\leq n$, we have the relations
\begin{equation}\label{bnalpha}
  B^{(n)}(\X_{l_1},\ldots,\X_{l_n})* \a_\m(\X_k) = B^{(n)}_{\ul{s_1,\cdots,s_n t_1 ,\cdots, t_n}} \bigg(H(\X_{k},s_{l^{-1}(k)},t_{l^{-1}(k)})\hspace{-5mm}\prod\limits_{j=1,j\neq l^{-1}(k)}^n \hspace{-5mm} C(\X_{l_j},s_j,t_j)\bigg)\,,
\end{equation}
\begin{equation}\label{alphabn}
 \a_\m(\X_k)*\, B^{(n)}(\X_{l_1},\ldots,\X_{l_n}) = B^{(n)}_{\ul{s_1,\cdots,s_n t_1 ,\cdots, t_n}} \bigg(E(\X_{k},s_{l^{-1}(k)},t_{l^{-1}(k)})\hspace{-5mm}\prod\limits_{j=1,j\neq {l^{-1}(k)}}^n \hspace{-5mm} C(\X_{l_j},s_j,t_j)\bigg)\,,
\end{equation}
where the formal notatin $l^{-1}(k)$ represents the index $j$ such that $l_j=k$. Notice that there are implicit $\a_\m^{st}$ dependence in the functions $E$ and $H$.

For $k>n$, the star product turns to the usual local commuting product
\begin{eqnarray}\label{bnalcmt}
 B^{(n)}(\X_1,\ldots,\X_n)* \a_\m(\X_k) &=&B^{(n)}(\X_1,\ldots,\X_n) \a_\m(\X_k) \\
\nn &=& \a_\m(\X_k)B^{(n)}(\X_1,\ldots,\X_n) =\a_\m(\X_k)* B^{(n)}(\X_1,\ldots,\X_n)\,,
\end{eqnarray}
which is zero after the projection \eqref{projection}.

We now consider the product
\begin{eqnarray}
   &&\hspace{-3mm}B^{(n)}(\X_{r_1},\ldots,\X_{r_n})* \a_\m(\X_k) *B^{(m)}(\X_{u_1},\ldots,\X_{u_m}) \label{bnalphabm}\\
 \nn &=&\hspace{-3mm}B^{(n)}_{\ul{s_1,\cdots,s_n t_1 ,\cdots, t_n}}B^{(m)}_{\ul{p_1,\cdots,p_m q_1 ,\cdots, q_m}} \bigg(H(\X_k,s_{r^{-1}k},t_{r^{-1}k})\hspace{-3mm}\prod\limits_{j=1,j\neq l^{-1}k}^n \hspace{-3mm} C(\X_{r_j},s_j,t_j)\bigg)*\prod\limits_{l=1}^m C(\X_{u_l},p_l,q_l)\,.
\end{eqnarray}
To evaluate the star product in the above expression,
notice that there are three possibilities
\begin{enumerate}
  \item  $\{\X_{\ul{r}}\} \bigcap\, \{\X_{\ul{u}}\}= \emptyset$,  the star product reduces to the usual commuting product. This reduces to zero after the projection \eqref{projection} since there are either $\bar{Y}$ or $\bar{Z}$ in the normal products.\fn{Note that here we use the fact that $|s|\neq 0$ and $|t|\neq 0$.}
  \item $\{\X_{\ul{r}}\} \bigcap\, \{\X_{\ul{u}}\}= \{\X_k\}$, which means $\X_k$ is the only element of $\{\X_{\ul{r}}\} \bigcap\, \{\X_{\ul{u}}\}$. In this case, the only non-trivial star products is between the $H$ function with the $C$ function that is in the same species of $H$.
  \item other cases, namely $\{\X_{\ul{r}}\} \bigcap\, \{\X_{\ul{u}}\}\supsetneq \{\X_k\}$ or $\{\X_{\ul{r}}\} \bigcap\, \{\X_{\ul{u}}\}\nsupseteq \{\X_k\}$, the term evaluates to zero by \eqref{cidty}.
\end{enumerate}
Therefore expression \eqref{bnalphabm} is zero unless $\{\X_r\} \bigcap\, \{\X_l\}=\{\X_k\}$, in which case \eqref{bnalphabm} evaluates to
\begin{eqnarray}
 \nn  &&B^{(n)}(\X_{r_1},\ldots,\X_{r_n})* \a_\m(\X_k) *B^{(m)}(\X_{u_1},\ldots,\X_{u_m}) \\
\nn &=&-r^{-2}\, B^{(n)}_{\ul{s_1,\cdots,s_n t_1 ,\cdots, t_n}}\a_\m^{\underline{t_{r^{-1} k}\,  p_{u^{-1}k}}}B^{(m)}_{\ul{p_1,\cdots,p_m q_1 ,\cdots, q_m}}\\
\nn&&\qquad \times C(\X_{k},s_{r^{-1}k},q_{u^{-1}k}) \hspace{-5mm} \prod\limits_{j=1,j\neq r^{-1}k}^n \hspace{-5mm} C(\X_{r_j},s_j,t_j)\hspace{-5mm}\prod\limits_{l=1,l\neq u^{-1}k}^m \hspace{-5mm} C(\X_{u_l},p_l,q_l)\\
&=& -r^{-2}\,
B^{(n)}_{\ul{s_1,\cdots,s_n t_1 ,\cdots, t_n}}\a_\m^{\underline{t_{r^{-1} k}\,  p_{u^{-1}k}}} B^{(m)}_{\ul{p_1,\cdots,p_m q_1 ,\cdots, q_m}}  \hspace{-5mm}\prod\limits_{j=\text{free indices}} \hspace{-6mm} C(\X_j,s_j(p_j),t_j(q_j))\,,\label{bnalphabmrstfn}
\end{eqnarray}
where the sum over repeated indices is implicit.

With all these relations, we are ready to consider the following definition of the gauge connection
\begin{eqnarray}\label{wnconnection}
  && B=\sum\limits_{i}B^{(1)}(\X_{i})+\frac{1}{n \, N^{n-1}}\sum\limits_{j_1,..,j_n}B^{(n)}
(\X_{j_1},\ldots,\X_{j_n})\label{wnconnectionb}\\
&&W^{(0)}_{r}=\frac{1}{r}\sum\limits_{j=1}^n\bigg(\bar{(z_j)}_{r} (z_j)^{\bullet}-\bar{(z_j)}_{\bullet}(z_j)^{r}\bigg)\\
&&W^{(0)}_a=\frac{1}{r}\sum\limits_{j=1}^n\bigg(\bar{(z_j)}_a\,
\big((z_j)^{\bullet}-(z_j)^{r}\big)-\big(\bar{(z_j)}_{\bullet}-
\bar{(z_j)}_{r}\big)\,(z_j)^a\bigg)\label{wnconnectionw0}\\
&& \widetilde{W}=W^{(0)}\,, \qquad W=W^{(0)}+\del W\,, \quad \del W= B* \sum\limits_{k} \a_\m(\X_k)\,,\label{wnconnectionwall}
  \end{eqnarray}
where we have introduced a formal sum over all possible flavor indices $i,j,k$. This makes both the $B$ and $W_r$ field symmetric under any permutation of the species indices $i$ and $j$, as required by the multiparticle algebra.\fn{Notice that this symmetric property can also be seen in the field theory side, which corresponds to the permutations of all the pairs $(p_1,q_i)$ in \eqref{sgdeformUV}.}

Then a straightforward but cumbersome computation shows that
\begin{eqnarray}
 &&\hspace{-8mm} (d_r\,B+W_r*B-B*\widetilde{W}_r)\, r^2 \\
\nn &&\hspace{-8mm}=  \sum\limits_{\ul{s},\ul{t}}\bigg[d_\L B_{\underline{st}}+B_{\underline{sx}}\,
  \a^{\underline{xy}}\,B_{\underline{yt}}-\L^{-1}(|s|+|t|+D-2)\,
  B_{\underline{st}}\bigg]
  \sum\limits_j C(\X_j,s,t)\\
\nn  &\hspace{-15mm}&\hspace{-3mm}+\frac{1}{n\,N^{n-1}}\sum\limits_{\s\in S_N} \sum\limits_{\ul{s_i},\ul{t_i}} \bigg[d_\L B^{(n)}_{\ul{s_1},...\ul{s_n}, \ul{t_1},...\ul{t_n}}+\L^{-1} (n(2-D)-\sum_i(|\ul{s_i}|+|\ul{t_i}|))B^{(n)}_{\ul{s_1},...\ul{s_n}, \ul{t_1},...\ul{t_n}}\\
\nn &\hspace{-15mm}&\hspace{-3mm}-\sum\limits^n_{k=1} \big( (B^{(n)}_{\ul{s_1},...\ul{s_n}, \ul{t_1},...\ul{t_n}}\big|_{\ul{t_k}=\ul{a}})\,\a^{\ul{a},\ul{b}}B^{(1)}_{\ul{b},t_k}+ B^{(1)}_{\ul{s_k},\ul{a}}\,\a^{\ul{a},\ul{b}}(B^{(n)}_{\ul{s_1},...\ul{s_n}, \ul{t_1},...\ul{t_n}}\big|_{\ul{s_k}=\ul{b}})\big)\bigg]\prod\limits_{j=1}^n C(\X_{\s(j)},s_j,t_j)\\
\nn  &&\hspace{-3mm}+\frac{1}{n^2\, N^{2n-2}}\hspace{-1mm}\sum\limits_{\s,\s'\in S_N}\hspace{-1mm}\bigg(\sum\limits^n_{k=1}\sum\limits_{\ul{s_i},\ul{t_i},\ul{u_i},\ul{v_i}} \hspace{-2mm}B^{(n)}_{\ul{s_1},...\ul{s_n}, \ul{t_1},...\ul{t_n}}\big|_{\ul{t_k}=\ul{a}}\,\a^{\ul{a},\ul{b}}B^{(n)}_{\ul{u_1},...\ul{u_n}, \ul{v_1},...\ul{v_n}}\big|_{\ul{u_k}=\ul{b}}\bigg)\\
&& \qquad\qquad \times\prod\limits_{ j=1}^{n}  C(\X_{\s(j)},s_j,t_j)C(\X_{\s'(j)},u_j,v_j)\,,\label{bwr}
\end{eqnarray}
where we have performed the projection \eqref{projection} when going from the first line to the second line.
The terms in the two square brackets vanish due to the ERG equations \eqref{ergfnl} and \eqref{nptrun} on the boundary. Thus we have
\begin{eqnarray}
\nn &&\hspace{-8mm} (d_rB+W_r*B-B*\widetilde{W}_r)\, r^2 = \\
\nn &&\hspace{-8mm}\frac{1}{n^2\, N^{2n-2}}\hspace{-3mm}\sum\limits_{\s,\s'\in S_N}\hspace{-3mm}\bigg(\sum\limits^n_{k=1}
\hspace{-1mm}B^{(n)}_{\ul{s_1},...\ul{s_n}, \ul{t_1},...\ul{t_n}}\big|_{\ul{t_k}=\ul{a}}\,\a^{\ul{a},\ul{b}}B^{(n)}_{\ul{u_1},...\ul{u_n}, \ul{v_1},...\ul{v_n}}\big|_{\ul{u_k}=\ul{b}}\bigg) \prod\limits_{ j=1}^{n}  C(\X_{\s(j)},s_j,t_j)C(\X_{\s'(j)},u_j,v_j)\\
&&\qquad \Rightarrow d_rB+W_r*B-B*W_r \simeq 0\,, \label{wblargen}
\end{eqnarray}
Notice that in our setup $n\geq 3$, the non-vanishing terms are significantly suppressed in the large-$N$ limit. We have used the sign $\simeq$ to represent the fact that \eqref{wblargen} is valid only in the large-$N$ limit. In this sense, our mapping to the bulk DMR higher-spin theory is only exact in the large-$N$ limit: the equation of motion will be deformed by $o(N)$ terms at finite $N$.

The computation in the $a\neq r$ direction follows in parallel. It is easy to generalize the result \eqref{Bifinal} to the interacting theory with interaction  $B^{(n)}(\f\bar\f)^n$, where the translation invariance in $x_a$ directions requires
\begin{eqnarray}
  d_{e^a} B^{(n)}_{\ul{s_1},...\ul{s_n}, \ul{t_1},...\ul{t_n}}\prod\limits_{i=1}^n p_i^{\ul{s_i}}q_i^{\ul{t_i}} \hspace{-2mm}&=&\hspace{-2mm} -i \sum\limits_{j=1}^n(p_j^a-q_j^a)B^{(n)}_{\ul{s_1},...\ul{s_n}, \ul{t_1},...\ul{t_n}}\prod\limits_{i=1}^n p_i^{\ul{s_i}}q_i^{\ul{t_i}}\label{nptspace}\\
\nn &&\hspace{-30mm}+\sum\limits^n_{k=1} \bigg( (B^{(n)}_{\ul{s_1},...\ul{s_n}, \ul{t_1},...\ul{t_n}}\big|_{\ul{t_k}=\ul{a}})\,\a^{\ul{a},\ul{b}}B^{(1)}_{\ul{b},t_k}+ B^{(1)}_{\ul{s_k},\ul{a}}\,\a^{\ul{a},\ul{b}}(B^{(n)}_{\ul{s_1},...\ul{s_n}, \ul{t_1},...\ul{t_n}}\big|_{\ul{s_k}=\ul{b}})\bigg)\prod\limits_{i=1}^n p_i^{\ul{s_i}}q_i^{\ul{t_i}}\,.
\end{eqnarray}
In the above computation in the free case, we have shown that the linear terms in \eqref{nptspace} are reproduced by $[W_a^{(0)},B]_*$ with the definition \eqref{wnconnectionb}, \eqref{wnconnectionw0} and \eqref{wnconnectionwall}. In addition the nonlinear terms in \eqref{nptspace} are reproduced by the $\del W_a*B=r\, B*\a_a*B$ term, which is identical to the computation in $r$ component as in \eqref{bwr} and \eqref{wblargen}. This is again due to the uniformality of the analogous expressions of \eqref{b1run} and \eqref{b1runspace1}  for the interacting theories. In summary, we can follow the same path of the derivation of \eqref{wblargen} to show that
\begin{equation}\label{bintspace}
   d_a B+W_a*B-B*W_a \simeq 0\,.
\end{equation}

With \eqref{wblargen} and \eqref{bintspace}, we can follow the same argument in \cite{Douglas:2010rc}, which is reviewed at the end of section \ref{mapfreesec} to show the following
\begin{equation}\label{wwlargen}
  dW+W*W\simeq0\,,\qquad  d\widetilde{W}+\widetilde{W}*\widetilde{W}\simeq0\,,
\end{equation}
where the $\simeq$ again represents the fact that the equations are true only in the large-$N$ limit.

\subsection{Deforming with more general interactions}
In this section, we consider the action with general interactions, that is, the interacting part of the action admits an expansion
\begin{equation}\label{gensint}
  S_{\text int}=\sum\limits_{n\geq 3} \frac{1}{N^{n-1}}B^{(n)}(p_1,\ldots,p_n,q_1,\ldots,q_n)\f^{A_1}({q_1}
)\bar{\f}^{A_1}({p_1})\cdots\f^{A_n}({q_n})
\bar{\f}^{A_n}{(p_n)}\end{equation}
We can directly compute the ERG of all the coupling constants $B^{(n)}$. Since all higher order interactions are included in the initial lagrangian \eqref{gensint}, the approximation we made in deriving \eqref{nptrun1} is no longer valid: we can no longer neglect the $ B^{(n+1)}+\sum_{l=2}^n B^{(l)} B^{(n-l+1)}$ term in the $d_\L B^{(n)}$ ERG equation. Therefore the ERG equations of the generic interacting lagrangian \eqref{gensint} take the following schematic form
\begin{equation}\label{erggensch}
  d_\L B^{(n)}= B^{(n+1)}+\sum_{l=1}^n B^{(l)} B^{(n-l+1)}\,,\quad n\geq 3\,.
\end{equation}
To map these equations to some higher-spin equations of motion, we generalize the construction in the previous section straightforwardly. Consider
\begin{eqnarray}\label{genb}
    && B=\sum\limits_{i}B^{(1)}(\X_{i})+\sum\limits_{n\geq 3}\frac{1}{n \, N^{n-1}}\sum\limits_{j_1,..,j_n}B^{(n)}
(\X_{j_1},\ldots,\X_{j_n})\label{wnconnectionb}\\
&&W^{(0)}_{r}=\frac{1}{r}\sum\limits_{j=1}\bigg(\bar{(z_j)}_{r} (z_j)^{\bullet}-\bar{(z_j)}_{\bullet}(z_j)^{r}\bigg)\\
&&W^{(0)}_a=\frac{1}{r}\sum\limits_{j=1}\bigg(\bar{(z_j)}_a\,
\big((z_j)^{\bullet}-(z_j)^{r}\big)-\big(\bar{(z_j)}_{\bullet}-
\bar{(z_j)}_{r}\big)\,(z_j)^a\bigg)\label{wnconnectionw0}\\
&& \widetilde{W}=W^{(0)}\,, \qquad W=W^{(0)}+\del W\,, \quad \del W= B* \sum\limits_{k} \a_\m(\X_k)\,,\label{wnconnectionwall}
\end{eqnarray}
Then a straightforward computation shows that the equation $d_r\,B+W_r*B-B*\widetilde{W}_r$ captures all the terms of the schematic form $B^{(l)} B^{(n-l+1)}$ but not the terms of the form $ B^{(n+1)}$, which comes from the $\frac{\pa^2 S_{\rm int}}{\pa \bar{\f}(p) \pa {\f}(p)}$ term in \eqref{erg}. This mismatch is cured in the large $N$ limit where all $B^{(n+1)}$ terms are suppressed by their $1/N^{n-1}$ coefficients, which show up in exactly the same way as the  $1/N^{n-1}$ coefficients in the second line of \eqref{bwr}. The only exception is the term with $n=1$, where the term $B^{(2)}$ in the $d_\L B^{(1)}$ equation, \eqref{erggensch}, is not suppressed with power of $1/N$. For this special case, we use the fact that in \eqref{gensint} we do not include the relevant deformation $B^{(2)}$, so the $B^{(2)}$ term appearing  in the $d_\L B^{(1)}$ equation must be generated dynamically along the RG flow and hence is of order $\co(\e)$ by our discussion at the beginning of this section. Therefore, it is subleading too.

In summary, for the most generic irrelevant deformation \eqref{gensint}, namely provided that we do not include the $n=2$ term, we can always map the ERG equation describing the running  of \eqref{gensint} to the DMR type higher-spin equation of motion in the large-$N$ limit. When finite $N$ corrections are included, the curvature will in general be non-vanishing, which can be understood as $o({1/N})$ breaking of the exact higher-spin symmetry by interactions in agreement with the conclusion reached in \cite{Maldacena:2011jn, Maldacena:2012sf}.
\section{Conclusions}\label{Conclusions}
In this paper we have examined how the ERG equations of interacting vector models map to a  multiparticle higher-spin equation of motion. We hope, by developing this particular example, to have shed some light on the bigger question of the holographic interpretation of the RG equations. A few ingredients in our derivation play prominent r\^oles and they will probably be central in more general situations:

\begin{itemize}

\item Multi-trace deformations, typical interactions in the field theory, require an understanding of multi-particle states in the bulk.

\item Our result is only exact in the large $N$ limit. In general, there will be extra terms in the ERG equation \eqref{bwr} and the curvature in the higher-spin equation of motion, \eqref{wwlargen}, is non-zero. This reflects the fact that the introduction of interactions in the vector model breaks the higher-spin symmetry in complete agreement with the results of Maldacena and Zhiboedov \cite{Maldacena:2012sf}.

\item In this paper, we want to construct the duality relation between the free vector model and a bulk higher-spin theory. Therefore it is crucial for us to consider RG flows near the UV fixed point (the free theory) instead of RG flows from the UV theory to another IR theory. This is the reason why we have excluded the relevant deformation and
    only run down infinitesimally from the UV scale:
    otherwise the RG flow drives the free theory to the critical $O(N)$ model which is a stable IR fixed point.

    Notice that the relevant deformation is used in \cite{Hartman:2006dy, Giombi:2011ya} to show that the duality between Vasiliev¡¯s theory and the critical $O(N)$ model follows, order by order in $1/N$, from the duality with free field theory on the boundary. This fact and our analysis are thus complementary to each other in the sense of whether relevant or irrelevant deformations are considered.

\item We have introduced a projection \eqref{projection} in our computation, which is a crucial step in the mapping from the ERG equations to higher-spin equations of motion. This is similar to the projection used in the Vasiliev higher-spin theory which selects  the physical information.

\item Dimensionality plays an important role, as is expected in the standard treatment of RG flows. In the case of the free limit discussed in \cite{Douglas:2010rc} dimensionality was very secondary. We find that the inclusion of interactions modifies the map to the HS equations present in the free case by dimension-dependent structures. It is plausible that the general focusing of trajectories in RG space is translated to a structure of deformation in the higher-spin equations of motion. This is a topic worth pursuing in more detail.

\item Having mapped ERG equations to higher-spin equations we see an explicit example, albeit simplified, of the ``RG=GR'' equation where GR is really a higher-spin gauge theory. By using the higher-spin formalism for the connection we explicitly answer the question of how the RG equations which are first order are equivalent, in the appropriate limit, to the Einstein equations which are second order in the metric formulation but first order in the connection formulation. More importantly, we have taken a step towards a ``covariantization'' or the ERG equations \eqref{erg},\eqref{ergspace} which is needed to make full contact with any covariant gravity theory.

\end{itemize}

Much more is known about vector models than their RG equations \cite{Moshe:2003xn}. It would be interesting to map other structures of the vector models on the boundary to the bulk higher-spin theory. In particular, the gap equation which has recently been discussed in the context of the higher-spin/vector model duality \cite{Leigh:2012mz}, should find a conceptual place within our formalism.  It will also be interesting to pursue what the identification discussed in this paper and its implications can teach us about quantum aspects of higher-spin gauge theories and their relations to full blown string theories. Another direction is to generalize the current discussions to lower dimensions, in particular, to $AdS_3/CFT_2$ holography with higher-spin fields where remarkable progress has been made recently \cite{Gaberdiel:2010pz,Gaberdiel:2011wb,Gaberdiel:2011zw,
Creutzig:2011fe,Chang:2011vka,Candu:2012jq,Gaberdiel:2012ku,
Gaberdiel:2012uj,Candu:2012tr,Creutzig:2012ar,Candu:2012ne,Peng:2012ae,
Chang:2013izp}.

\section*{Acknowledgments}
L. PZ thanks Rob Leigh, Djordje Minic and Diana Vaman for collaboration on related topics and the Aspen Center for Physics for hospitality. CP is grateful for helpful discussions with Fang Chen, Nabil Iqbal, Joseph Polchinski and Douglas Stanford. We thank  Henriette Elvang, Matthias Gaberdiel, Thomas Hartman, Yu-tin Huang, Shlomo Razamat and Dori Reichmann for their useful comments on this project and on an early draft of this paper. This research was supported in part by the National Science Foundation under grant No. 1066293 (Aspen) and by the Department of Energy under grant DE-FG02-95ER40899 to the University of Michigan. CP is supported by NSF Grant PHY-0953232 and in part by the DoE Grant DE-SC0007859. He would like to thank the Kavli Institute for Theoretical Physics for the support as a KITP graduate fellow, and his work at KITP was supported in part by the National Science Foundation under Grant No. NSF PHY11-25915.

\bibliographystyle{JHEP}
\bibliography{RGHS}

\end{document}